\documentclass[aps,prd,showpacs,preprintnumbers,nofootinbib]{revtex4}
\usepackage[dvips]{graphicx}
\usepackage{bm,latexsym,amsmath,amssymb,amsfonts}
\begin{document}

\title{On asymptotic structure at null infinity in five dimensions}

\author{Kentaro Tanabe}
\affiliation{Yukawa Institute for Theoretical Physics, Kyoto University, Kyoto 606-8502, Japan}
\author{Norihiro Tanahashi and Tetsuya Shiromizu}
\affiliation{Department of Physics, Kyoto University, Kyoto 606-8502, Japan}
\begin{abstract}
We discuss the asymptotic structure of null infinity in five dimensional space-times. Since 
it is known that the conformal infinity is not useful for odd higher dimensions, we shall employ 
the coordinate based method such as the Bondi coordinate first introduced in four dimensions. 
Then we will define the null infinity and identify the asymptotic symmetry. We will also derive the 
Bondi mass expression and show its conservation law.  
\end{abstract}
\maketitle

\section{Introduction}

Inspired by superstring theory, fundamental studies of higher dimensional space-time is important. 
One issue is the 
asymptotic structure at null infinity. We often introduce the conformal infinity 
to discuss the asymptotic structure at infinities in four dimensions \cite{Penrose:1962ij, Wald}. 
Therein the 
space-time is compactified by conformal transformation and embedded into another 
space-time. For example, Minkowski space-time is conformally embedded into the Einstein 
static universe. Conformally embedded space-time has two different infinities, 
i.e., spatial infinity and null infinity. 
In higher dimensional space-times, 
the asymptotic structure at spatial infinity can be well-defined. 
The asymptotic symmetry is identified with the Poincare group and conserved quantities associated 
with the symmetry are constructed \cite{Shiromizu:2004jt, Tanabe:2009xb}. 

On the other hand, the asymptotic structure at null infinity is not completely understood 
 in higher dimensional space-times 
\cite{Hollands:2003ie, Hollands:2004ac, Ishibashi:2007kb}. Indeed, the definition of 
null infinity is given only in even dimensions \cite{Hollands:2003ie, Ishibashi:2007kb}. 
As seen later, the difficulty in defining null 
infinity as compared to spatial infinity is due to the presence of the gravitational 
waves at null infinity. Since there are no gravitational 
waves at spatial infinity, the asymptotic structure is ``stationary" and the total mass 
and total angular momentum are conserved. On the other hand, the asymptotic structure 
at null infinity might be disturbed by gravitational waves. Hence, we need a 
stable definition of null infinity against gravitational waves. 
%and this can be 
%done only in {\it even} dimensions by using conformal embedding. 
We can give such a definition in {\it even} dimensions if we use a conformal embedding method. 
But we cannot do this in odd dimensional space-times because we cannot show the smoothness of 
the Einstein equations at null infinity.
This non-smoothness would be related to the facts that in the conformal 
embedding method we introduce the conformal factor $\Omega \sim 1/r$, 
and the behavior of gravitational 
waves near null infinity are of the order ${\cal O}(\Omega^{(d-2)/2})$ in $d$ 
dimensional space-times. 
The problem comes from the half-integer power of $\Omega$.
%And
From this smoothness only, however, 
we cannot show that the boundary conditions at null infinity for asymptotic flatness 
in the papers~\cite{Hollands:2003ie, Hollands:2004ac, Ishibashi:2007kb} are the 
marginal and weakest conditions 
%whether there are gravitational waves at null infinity or not. 
which allow gravitational waves at null infinity to exist.
%Hence, to 
To find such marginal conditions, which the boundary conditions should be so, 
we must solve the Einstein equations near null infinity and clarify the freedom 
of gravitational waves.  
%-------------------------

In this paper, we define null infinity in five dimensional space-time. We do not use the
conformal embedding method, but instead we use the Bondi coordinate to define null infinity, which was 
introduced firstly by Bondi and Sachs in four dimensions \cite{Bondi, Wnicour, Sachs1, Sachs2}. 
The rest of this paper 
is organized as follows. In section 2, we introduce the Bondi coordinate in five 
dimensions. In section 3, we define the asymptotic flatness at null infinity in 
five dimensions and show the robustness of the definition against gravitational 
waves by solving Einstein equations. In this section, we define the Bondi mass, and obtain 
the mass loss law in five dimensions. In section 4, we discuss the asymptotic 
symmetry at null infinity associated with the asymptotic flatness defined in section 3. 
We show that there are no supertranslations in five dimensions unlike in four dimensions. 
Finally, in section 5, we give a discussion and summary.

\section{Bondi coordinate in five dimensions}

We consider five dimensional space-time. We introduce the Bondi coordinate to 
define asymptotic flatness at null infinity. 
Suppose there is a function $u(x^a)$ 
which satisfies the equation
%============<Equation>=============%
%
\begin{equation}
u_{,a}u_{,b} g^{ab} \,=\, 0,
\end{equation}
%
%===================================%
where Latin indices run from $0$ to $4$ and $u_{,a}=\partial u /\partial x^a$. 
Then $u=\text{constant}$ surfaces are null hypersurfaces, and we use this function $u$
as retarded time to construct a coordinate system. 
Let $\theta$, $\phi$, $\psi$ be angular coordinates, which are constant along gradient $u$. 
The period of each of these coordinates are taken to be $\pi$, $2\pi$, $2\pi$, respectively.
For convenience, we introduce the notation $( \theta, \phi, \psi ) = x^{A}$, 
where capital Latin indices run from $2$ to $4$. 
Now we define the function $r$ by the equation
%============<Equation>=============%
%
\begin{equation}
r^6 \sin^{2}\theta \cos^{2}\theta \,=\, {\rm det}(g_{AB}). \label{r-def}
\end{equation}
%
%===================================%

Using these coordinates (we call them the Bondi coordinates)
%============<Equation>=============%
%
\begin{equation}
x^{0}=u\, ,\,x^{1}=r\,,\,x^{2}=\theta\,,\,x^{3}=\phi \,,\,x^{4}=\psi ,
\end{equation}
%
%===================================%
we can write down metrics as 
%============<Equation>=============%
%
\begin{equation}
ds^{2}\,=\,-(V e^{B}/r^2) du^{2} - 2e^{B} du dr + r^{2}h_{AB}(dx^{A} + U^{A}du)(dx^{B} + U^{B}du), \label{metric}
\end{equation}
%
%===================================%
where
%============<Equation>=============%
%
\begin{equation}
h_{AB} \,=\, 
\begin{pmatrix}
e^{C_1} & \sin\theta \sinh D_1 & \cos\theta \sinh D_2 \\
\sin\theta \sinh D_{1} & e^{C_2} \sin^{2} \theta & \sin\theta \cos\theta \sinh D_3 \\
\cos\theta \sinh D_2 & \sin\theta \cos\theta \sinh D_3 & e^{C_3} \cos^{2} \theta
\end{pmatrix} .
\end{equation}
%
%===================================%
In the above, $V, B, h_{AB}, U^A, C_1, C_2, C_3, D_1, D_2$ and $D_3$ are functions of $u, r$ and $x^A$. 
In this coordinate, null infinity is represented by 
$r\rightarrow\infty$.

From Eq.~(\ref{r-def}), we have a relation between 
$C_1, C_2, C_3, D_1, D_2$ and $D_3$ as 
%============<Equation>=============%
%
\begin{equation}
e^{C_3} \,=\, \frac{1 + e^{C_2} \sinh^{2} D_2 + e^{C_1}\sinh^{2}D_3 -2\sinh D_1 \sinh D_2 \sinh D_3 }{e^{C_1+C_2} - \sinh^{2}D_1} .  \label{C3-relation}
\end{equation}
%
%===================================%
As we will realize later, these five independent functions $C_1, C_2, D_1, D_2$ and $D_3$ 
correspond to the degrees of freedom of gravitational waves in five dimensional space-times.

\section{Einstein equation at null infinity}
As stated in the introduction, the definition of null infinity should be not disturbed by  
gravitational waves. The robustness of the null infinity definition implies that the boundary conditions 
imposed on the metric (\ref{metric}) should be compatible with Einstein equations. 
Hence, in order to define asymptotic structure at null infinity and show the robustness of this definition, 
we should specify the proper boundary conditions by solving Einstein equations in 
the Bondi coordinate.  

\subsection{Vacuum Einstein equations}

%In the current set-up, for convenience, we decompose 
%the vacuum Einstein equation into trivial equation
%$R_{ur}=0$, constraint equations $
%R_{rr}\,=\,0\,,\,R_{rA}\,=\,0\,,\,R_{AB}\,=\,0
%$
%and
%evolution equations
%$
%R_{uu}\,=\,0\,,\,R_{uA}\,=\,0.
%$
%Roughly speaking, trivial equation and evolution equations describe 
%impose constraint conditions on the metric of such surfaces. From now on, 
%we will examine each equations, respectively. 

Since equations are very complicated, we will not write down explicit forms here. 
We will show the equations in {\it a symbolic way} in order to see only the essential 
structure of equations. See Appendix A for some details of the equations which will be 
used. 

%\subsubsection{Constraint equation}

From $R_{rr}=0$, we have 
%============<Equation>=============%
%
\begin{equation}
\frac{\partial B}{\partial r} \,=\, \frac{\partial \mathbb{C}^{2}}{\partial r}, \label{B-rel}
\end{equation}
%
%===================================%
where $\mathbb{C}$ stands for $C_1, C_2, D_1, D_2$ and $D_3$. From $R_{rA}=0$, 
%============<Equation>=============%
%
\begin{equation}
\frac{\partial}{\partial r} \left( r^5 \frac{\partial U^{A}}{\partial r} \right) = r^2 \left(  \mathbb{C}  + \mathbb{C}^{2} + \cdots \right) \label{U-rel}
\end{equation}
%
%===================================%
The trace and traceless part of $R_{AB}=0$ implies 
%============<Equation>=============%
%
\begin{eqnarray}
\frac{\partial} {\partial r}( r^2 e^{-B}V) = \eta (\mathbb{C},U^A) \label{V-rel}
\end{eqnarray}
%
%===================================%
and
%============<Equation>=============%
%
\begin{eqnarray}
\frac{\partial^{2}}{\partial u\partial r} \mathbb{C} 
=\delta (\mathbb{C},U^A), \label{C-rel}
\end{eqnarray}
%
%===================================%
where $\eta$ and $\delta$ are some functions of $\mathbb{C}$ and $U^A$. 

Since we integrate the equations with respect 
to $r$ in solving  the evolution equations, 
some integration functions $f(u,x^{A})$ appear.  
Constraint equations describe the evolution of such functions.

In the Bondi coordinates, after integrating the other equations, we can show that  
the equation $R_{ur}=0$ would be satisfied trivially. 
%This is because $r=\text{constant}$ surface is defined rigidly 
%by the condition (\ref{r-def}) on any $u$. 

%\subsubsection{Evolution equations}

The evolution equations $R_{uu}=0$ and $R_{uA}=0$ have the following form 
%============<Equation>=============%
%
\begin{equation}
\frac{\partial V}{\partial u} \,=\, r^3( \mathbb{C}^{2} + \mathbb{C}^{3} +\cdots ) + r^{2} ( \mathbb{C} + \mathbb{C}^{2} + \cdots ) \label{M-evo}
\end{equation}
%
%===================================%
and 
%============<Equation>=============%
%
\begin{equation}
r\frac{\partial U^{A}}{\partial u} \,=\, \frac{\partial \mathbb{C}}{\partial u}, \label{J-evo}
\end{equation}
%
%===================================%
respectively. 

If $\mathbb{C}$ is given on initial surface $u=u_{0}$, we can determine the metric function 
$B$, $U^{A}$,$V$ from Eqs.~(\ref{B-rel}), (\ref{U-rel}) and (\ref{V-rel}) except 
for integration functions. The evolution of $\mathbb{C}$ and integration functions are 
described by Eqs.~(\ref{C-rel}), (\ref{M-evo}) and (\ref{J-evo}). 

As seen later,  
the functions $\mathbb{C}=(C_1,C_2,D_1,D_2,D_3)$ will be identified with the freedom of gravitational waves, 
and Eq.~(\ref{M-evo}) will govern the evolution of their total mass.
Thus, to obtain a stable definition of null infinity, we should determine the asymptotic 
behavior of the function $\mathbb{C}$. Then, using Eqs.~(\ref{B-rel}), (\ref{U-rel}) and (\ref{V-rel}), 
we can obtain the asymptotic behavior of $B$, $U^{A}$, $V$, and the robustness against gravitational 
waves of these boundary conditions would be guaranteed by the evolution equation (\ref{M-evo})
and Eq. (\ref{J-evo}).  

\subsection{Asymptotic behavior of gravitational fields}

From now on, we will write down all equations explicitly. 
When functions $\mathbb{C}$ corresponds to gravitational waves, 
$\mathbb{C}$ behaves as $\sim 1/r^{3/2}$ 
near null infinity. 
This can been seen from the solutions to the wave equation and/or 
the finiteness of the gravitational flux at null infinity. Therefore, we assume 
the behavior of $C_1, C_2, C_3, D_1, D_2$ and $D_3$ near 
null infinity such that
%============<Equation>=============%
%
\begin{gather}
C_1(u,r,x^{A})\,=\,\frac{C_{11}(u,x^{A})}{r^{3/2}} +O(1/r^2)  \\
C_2(u,r,x^{A})\,=\,\frac{C_{21}(u,x^{A})}{r^{3/2}} +O(1/r^2)  \\
C_3(u,r,x^{A})\,=\,\frac{C_{31}(u,x^{A})}{r^{3/2}} +O(1/r^2)  \\
D_1(u,r,x^{A})\,=\,\frac{D_{11}(u,x^{A})}{r^{3/2}} +O(1/r^2)  \\
D_2(u,r,x^{A})\,=\,\frac{D_{21}(u,x^{A})}{r^{3/2}} +O(1/r^2)  \\
D_3(u,r,x^{A})\,=\,\frac{D_{31}(u,x^{A})}{r^{3/2}} +O(1/r^2).  
\end{gather} 
%
%===================================%
As noted around Eq.~(\ref{C3-relation}), $C_{31}$ can be written as
%============<Equation>=============%
%
\begin{equation}
C_{31}=-(C_{11}+C_{21}).
\end{equation}
%
%===================================%
Then, from Eq.~(\ref{B-rel}), we see
%============<Equation>=============%
%
\begin{equation}
B= B_1/r^3 +O(r^{-7/2})
\end{equation}
%
%===================================%
near null infinity, and we obtain a relation
%============<Equation>=============%
%
\begin{equation}
B_1=-\frac{(C_{11}^{2}+C_{11}C_{21}+C_{21}^{2}+D_{11}^{2}+D_{21}^{2}+D_{31}^{2})}{8}.
\end{equation}
%
%===================================%
Furthermore, integrating Eqs.~(\ref{U-rel}), we have the relations
%============<Equation>=============%
%
\begin{eqnarray}
U_{\theta 1}=\frac{2(C_{11}\cot\theta - C_{21}\cot\theta - 2C_{11}\tan\theta 
- C_{21}\tan\theta + C_{11,\theta}
 + D_{11,\phi}\csc\theta + D_{21,\psi}\sec\theta )}{5},
\end{eqnarray}
%
%===================================%
%============<Equation>=============%
%
\begin{eqnarray}
U_{\phi 1}\sin\theta =\frac{2( 2D_{11}\cot\theta - D_{11}\tan\theta + D_{11,\theta} 
+C_{21,\phi}\csc\theta +D_{31,\psi}\sec\theta )}{5} 
\end{eqnarray}
%
%===================================%
and
%============<Equation>=============%
%
\begin{eqnarray}
U_{\psi 1}\cos\theta = \frac{2( D_{21}\cot\theta - 2D_{21}\tan\theta + D_{21,\theta} 
+ D_{31,\phi}\csc\theta
-(C_{11,\psi} + C_{21,\psi})\sec\theta )}{5} ,
\end{eqnarray}
%
%===================================%
where $U_{A1}$ are coefficients of the expansions defined as 
%============<Equation>=============%
%
\begin{equation}
U^{A} = \frac{U_{A1}}{r^{5/2}} +O(1/r^3). 
\end{equation}
%
%===================================%
Then, from Eq.~(\ref{V-rel}), we find
%============<Equation>=============%
%
\begin{equation}
V\,=\,r^{2} +V_1(u,x^A) r^{1/2} - M(u,x^A) +O(r^{-1/2}),
\end{equation} 
%
%===================================%
where
%============<Equation>=============%
%
\begin{equation}
V_1\,=\,-\frac{2}{3}\left( \frac{1}{\sin\theta \cos\theta}(U_{\theta 1} \sin\theta \cos\theta)_{,\theta}
 + U_{\phi 1,\phi} +U_{\psi 1,\psi}\right)
\end{equation}
%
%===================================%
and $M(u,x^{A})$ is an integration constant.

In Eq.~(\ref{C-rel}), 
$C_{11,u}$, $C_{21,u}$, $D_{11,u}$, $D_{21,u}$ and $D_{31,u}$ do not appear and then 
this means that we can fix them freely as initial conditions. These freedom correspond to 
the degree of freedom of gravitational waves in five dimensions. 

%Finally, from Eq.~(\ref{V-rel}), we get $\partial_u M(u,x^{A})$ as  
Finally, from Eq.~(\ref{M-evo}), we obtain the following formula
%============<Equation>=============%
%
\begin{eqnarray}
\frac{\partial M (u,x^A)}{\partial u} \,&=&\,-\frac{1}{3}\Biggl( 
%C_{11}^{2} +C_{11}C_{21} + C_{21}^{2} +D_{11}^{2} + D_{21}^{2} +D_{31}^{2} 
(C_{11,u})^{2} +C_{11,u}C_{21,u} +(C_{21,u})^{2} +(D_{11,u})^{2}+(D_{21,u})^{2} +(D_{31,u})^{2}
\Biggr) \notag \\ 
&=&\,-\frac{1}{3}\Biggl(
(C_{11,u}+C_{21,u}/2)^{2} + 3(C_{21})^{2}/4 + (D_{11,u})^{2}+(D_{21,u})^{2} +(D_{31,u})^{2}
\Biggr) \label{massloss} .
\end{eqnarray}
%
%===================================%
Eq.~(\ref{massloss}) represents mass loss rate by gravitational waves, 
and total mass always decreases as in four dimensions. Then $M(u,x^A)$ describes the mass in 
$u=\text{constant}$ surfaces. 
 
\subsection{Boundary conditions}
As shown in the previous subsection, in the presence of gravitational waves, asymptotic behaviors 
of metric (\ref{metric}) in leading order should be 
%============<Equation>=============%
%
\begin{eqnarray}
& & V=r^{2} +V_{1}(u,x^{A})r^{1/2} - M(u,x^{A})+O(1/r^{1/2}) \label{bc1}\\
& & B= \frac{B_{1}(u,x^{A})}{r^3} + O(1/r^{7/2}) \label{bc2}\\
& & U^{A} = \frac{U_{A1}(u,x^{A})}{r^{5/2}} +O(1/r^{3}) \label{bc3}\\
& & \mathbb{C}=\frac{\mathbb{C}_{1}(u,x^{A})}{r^{3/2}} +O(1/r^2) \label{bc4} .
\end{eqnarray}
%
%===================================%
In four dimensions, boundary conditions at null infinity for asymptotic flatness are determined by 
these leading behavior. As shown below, however, in higher dimensions than four, further conditions are
needed for asymptotic flatness. 

We consider asymptotic behavior in next-to-leading order terms. 
The function $\mathbb{C}$ can be expanded as follows
%============<Equation>=============%
%
\begin{equation}
\mathbb{C}\,=\,\frac{\mathbb{C}_1(u,x^{A})}{r^{3/2}} 
+ \frac{\mathbb{A}(u,x^{A})}{r^{2}} +
\frac{\mathbb{C}_2(u,x^{A})}{r^{5/2}} + 
\frac{\mathbb{C}_3(u,x^{A})}{r^{3}} + O(1/r^{7/2}) .
\end{equation} 
%
%===================================%
The equations $R_{AB}=0$, which describes the evolution of $\mathbb{C}$,  becomes
%============<Equation>=============%
%
\begin{equation}
\frac{\mathbb{A}_{,u}}{r}+O(r^{-3/2})=0
\end{equation}
%
%===================================%
and then we see
%============<Equation>=============%
%
\begin{equation}
\frac{\partial \mathbb{A}}{\partial u} \,=\,0 \label{C_{2}-rel}.
\end{equation}
%
%===================================%

At spatial infinity, asymptotic flatness requires that the Weyl tensor on spatial infinity 
behave like $\sim r^{-5}$ in order for the Taub-NUT charge to vanish \cite{Tanabe:2009xb}. 
This implies that $\mathbb{A}$ vanishes at spatial infinity ($u=-\infty$). 
From Eq. (\ref{C_{2}-rel}), $\mathbb{A}$ should vanish for asymptotic flatness at null infinity 
%============<Equation>=============%
%
\begin{equation}
\mathbb{A}\,=\,0.
\end{equation}
%
%===================================%
Then, from Eqs.~(\ref{V-rel}), (\ref{B-rel}) and (\ref{U-rel}), we can show that 
other metric functions can be expanded as follows
\footnote{$U_{A3}$ terms correspond to angular momentum. This can be seen from the comparison with the 
asymptotic behavior of Myers-Perry solutions \cite{MP}. }
%============<Equation>=============%
%
\begin{eqnarray}
& & V=r^{2} +V_1(u,x^{A})r^{1/2} -M(u,x^{A}) +O(1/r^{1/2}), \label{bc6}\\
& & B=\frac{B_1(u,x^{A})}{r^{3}} + \frac{B_2(u,x^{A})}{r^{4}} + \frac{B_3(u,x^{A})}{r^{9/2}} +O(1/r^{5}), \label{bc7} \\
& & U^{A}=\frac{U_{A1}(u,x^{A})}{r^{5/2}}+\frac{U_{A2}(u,x^{A})}{r^{7/2}}+\frac{U_{A3}(u,x^{A})}{r^{4}}
+O(1/r^{9/2}) \label{bc8}.
\end{eqnarray}
%
%===================================%
Thus, boundary conditions at null infinity for asymptotic flatness are
\begin{equation}
\mathbb{C}\,=\,\frac{\mathbb{C}_1(u,x^{A})}{r^{3/2}} 
+\frac{\mathbb{C}_2(u,x^{A})}{r^{5/2}} + 
\frac{\mathbb{C}_3(u,x^{A})}{r^{3}} + O(1/r^{7/2}) \label{bc5},
\end{equation} 
and Eqs. (\ref{bc6}), (\ref{bc7}), (\ref{bc8}).

\section{Asymptotic symmetry at null infinity}

In this section, we consider asymptotic symmetry at null infinity. Asymptotic symmetry should 
be defined as transformations preserving the boundary conditions~(\ref{bc5}), 
(\ref{bc6}), (\ref{bc7}) and (\ref{bc8}). 
By infinitesimal transformation $\xi$, the metric is transformed as
%============<Equation>=============%
%
\begin{equation}
\delta g_{ab} \,=\, 2 \nabla_{(a}\xi_{b)}.
\end{equation}
%
%===================================%
To preserve the boundary conditions given in the 
previous section, the variation of metric, $\delta g_{ab}$, should satisfy
%============<Equation>=============%
%
\begin{gather}
\delta g_{rr}\,=\,0\,,\,\delta g_{rA}\,=\,0\,,\,g^{AB}\delta g_{AB}\,=\,0 ,\label{sym}\\
\delta g_{uu}\,=\,O(r^{-3/2})\,,\,\delta g_{ur}\,=\,O(r^{-3})\,,\,\delta g_{uA}\,=\,O(r^{-1/2})\,,\,\delta g_{AB}\,=\,O(r^{1/2}).
\label{sym2}
\end{gather}
%
%===================================%
Here, as a first step, we will consider leading order terms. 

From Eq.~(\ref{sym}), we see that the components of infinitesimal transformation $\xi$ 
must take the following forms: 
%============<Equation>=============%
%
\begin{eqnarray}
& & \xi_{r}=f(u,x^{A})e^{B} ,  \label{xir}\\
& & \xi_{B}g^{AB}=f^{A}(u,x^{A})-fU^{A}+\int^{\infty}_{r}dr'e^{B}f_{,B}g^{AB} ,  \label{xiA}\\
& & \xi_{u}=-\frac{r e^{B}}{3}\left(
-\xi_{A,B}+\xi_{C}\Gamma^{C}_{AB}+\xi_{r}\Gamma^{r}_{AB}
\right) g^{AB}\label{xiu}.
\end{eqnarray}
%
%===================================%

The infinitesimal transformation $\xi$ has four free parameters $f, f^{A}$. $f$ and 
$f^A$ corresponds to translations and Lorentz transformation, respectively.
The other components of metric variation become 
%============<Equation>=============%
\begin{eqnarray}
\delta g_{uu}&=&\frac{2r}{3}
%\frac{\partial}{\partial u}f^{A}_{~~;A} 
{\cal D}_A f^{A}_{~~,u} 
+\frac{2}{3}
(3f +{\cal D}^2 f)_{,u}+\frac{2}{r^{1/2}}
h^{(0)}_{AB}f^{A}_{~~,u}U_{B1} 
%f^{A}_{~~,u}U_{A1} 
+O(r^{-3/2}), \\
\delta g_{ur}&=&\frac{1}{3}({\cal D}_A f^{A}+3f_{,u}) 
+\frac{1}{5r^{5/2}}h^{(1)AB}{\cal D}_A {\cal D}_B f +O(r^{-3}), \\
\delta g_{uA}&=&r^{2}h^{(0)}_{AB}f^{B}_{~~,u} +\frac{r}{3}{\cal D}_A (3f_{,u} +{\cal D}_B f^{B}) 
+r^{1/2}h^{(1)}_{AB}f^{B}_{~~,u} 
%+\frac{1}{3}(f_{;B}^{~~B}+3f)_{,A}+O(r^{-1/2}), \\
+\frac{1}{3}{\cal D}_A ({\cal D}^2 f+3f)+O(r^{-1/2}), \\
\delta g_{AB}&=&\frac{2r^{2}}{3}(-{\cal D}_C f^{C}h^{(0)}_{AB} +3{\cal D}_{(A}f_{B)}) 
+\frac{2r}{3}(-{\cal D}^2fh^{(0)}_{AB}+3{\cal D}_A{\cal D}_B f) +T_{AB}(u,x^{A})r^{1/2} +O(r^{-1/2}),
\end{eqnarray}
%===================================%
where 
$X_{(AB)}:=(1/2)(X_{AB}+X_{BA})$ for some tensor 
$X_{AB}$ and $T_{AB}$ is some traceless tensor with respect to 
$h_{AB}^{(0)}$. In the above, we expanded the metric $h_{AB}$ as 
%============<Equation>=============%
\begin{equation}
h_{AB}\,=\,h^{(0)}_{AB}+\frac{1}{r^{3/2}}h^{(1)}_{AB} +O(r^{-5/2}),
\end{equation}
%===================================%
and
$
h^{(1)AB}=h^{(0)AC}h^{(0)BD}h^{(1)}_{CD}.
$
In the Bondi coordinate, $h^{(0)}_{AB}$ and $h^{(1)}_{AB}$ are
%============<Equation>=============%
%
\begin{equation}
h^{(0)}_{AB}=
\begin{pmatrix}
1 & 0 & 0 \\
0 & \sin^{2}\theta & 0 \\
0 & 0 & \cos^{2}\theta 
\end{pmatrix}
\end{equation}
%
%===================================%
and
%============<Equation>=============%
%
\begin{equation}
h^{(1)}_{AB}=
\begin{pmatrix}
C_{11} & D_{11}\sin\theta & D_{21}\cos\theta \\
D_{11}\sin\theta & C_{21}\sin^{2}\theta & D_{31}\sin\theta\cos\theta\\
D_{21}\cos\theta & D_{31}\sin\theta\cos\theta & -(C_{11}+C_{21})\cos^{2}\theta
\end{pmatrix}.
\end{equation}
%
%===================================%
Note that $h^{(1)}_{AB}$ is traceless, $h^{(0)AB}h^{(1)}_{AB}=0$. 

To satisfy the condition (\ref{sym2}), we find that $f^{A}$ and $f$ should satisfy the following 
equations:
%===========<Equation>==============%
%
\begin{eqnarray}
& & 
%\frac{\partial}{\partial u}f^{A}
f^{A}_{~~,u}
= 0  , \label{f0-cond}\\
& & {\cal D}_A f_B+{\cal D}_B f_A =-2\frac{\partial f}{\partial u}h_{AB}^{(0)}  ,  \label{fA-cond} \\
%& & f_{;AB}+f_{;BA}=\frac{2}{3}f_{;C}^{~~C}h_{AB}^{(0)}.\label{f-cond}
& & {\cal D}_A {\cal D}_B f=\frac{1}{3}{\cal D}^2fh_{AB}^{(0)}.\label{f-cond}
\end{eqnarray}
%
%===================================% 
Integrating the trace part of Eq.~(\ref{fA-cond}) under the condition (\ref{f0-cond}), we obtain
%===========<Equation>==============%
%
\begin{equation}
    f=-\frac{u}{3}F + \alpha(x^{A}), \label{f-fA}
\end{equation}
%
%===================================% 
where $F:={\cal D}_A f^{A}$ and 
$\alpha(x^{A})$ is an arbitrary function of $x^A$. 
Here we can show from Eq.~(\ref{fA-cond}) that $F$ satisfies 
%===========<Equation>==============%
%
\begin{equation}
{\cal D}_A{\cal D}_BF =\frac{1}{3}{\cal D}^2Fh^{(0)}_{AB}
\label{divfA}
\end{equation}
%
%===================================% 
and
%===========<Equation>==============%
%
\begin{equation}
{\cal D}^2F+3F=0. 
\label{divF}
\end{equation}
%
%===================================% 
See Appendix B for the derivation. 
The general solution to these is given by 
%===========<Equation>==============%
%
\begin{equation}
F=E_{x}\sin\theta\cos\phi 
+E_{y}\sin\theta\sin\phi +E_{z}\cos\theta\cos\psi +E_{w}\cos\theta\sin\psi,
\end{equation}
%
%===================================% 
where $E_{x},E_{y},E_{z},E_{w}$ are constants. Then,  
from Eqs. (\ref{f-cond}) and (\ref{f-fA}), we see that 
%===========<Equation>==============%
%
\begin{equation}
{\cal D}_A{\cal D}_B\alpha =\frac{1}{3}{\cal D}^2 \alpha h^{(0)}_{AB}
\label{alpha}
\end{equation}
%
%===================================% 
holds. The solution to this is 
%===========<Equation>==============%
%
\begin{equation}
\alpha=a_u+a_{x}\sin\theta\cos\phi 
+a_{y}\sin\theta\sin\phi +a_{z}\cos\theta\cos\psi +a_{w}\cos\theta\sin\psi,
\end{equation}
%
%===================================% 
where $a_u, a_{x},a_{y},a_{z},a_{w}$ are constants. 
As a summary the general solution for $f$ is given by
%Note that $f$ satisfying Eq.~(\ref{f-cond}) has five independent solution
%===========<Equation>==============%
%
\begin{equation}
f\,=\,e_{u} +e_{x}(u)\sin\theta\cos\phi 
+e_{y}(u)\sin\theta\sin\phi +e_{z}(u)\cos\theta\cos\psi 
+e_{w}(u)\cos\theta\sin\psi,
\end{equation} 
%
%===================================%
%where $e_{u},e_{x},e_{y},e_{z},e_{w}$ are constants, and such $f$ satisfies 
where $e_{u}:=a_u$ is constant, $e_{x}(u):=-(u/3)E_x+a_x, 
e_{y}(u):=-(u/3)E_y+a_y, e_{z}(u):=-(u/3)E_z+a_z$ and 
$e_{w}(u):=-(u/3)E_w+a_w$. Now we can show that 
%===========<Equation>==============%
%
\begin{equation}
{\cal D}_A ({\cal D}^2 f+3f)\,=\,0,
\qquad
\partial_u ({\cal D}^2f+3f)\,=\,0
\end{equation}
%
%===================================%
hold. 
Using the above equations, at a glance, we see that Eqs. (\ref{sym2}) become 
%============<Equation>=============%
\begin{eqnarray}
%\delta g_{uu}&=&\frac{1}{3}\frac{\partial}{\partial u}(3f+f_{;A}^{~~;A}) +O(r^{-3/2}), \\
\delta g_{uu}&=&O(r^{-3/2}), \label{dguu}\\
\delta g_{ur}&=&\frac{1}{5r^{5/2}}h^{(1)AB}{\cal D}_A{\cal D}_B f +O(r^{-3}), \label{dgur}\\
\delta g_{uA}&=&O(r^{-1/2}),\label{dguA} \\
\delta g_{AB}&=&\frac{2r^{2}}{3}(-{\cal D}_C f^{C}h^{(0)}_{AB} +3{\cal D}_{(A}f_{B)}) 
+T_{AB}(u,x^{A})r^{1/2} +O(r^{-1/2}). \label{dgAB}
\end{eqnarray}
%===================================%
$h^{(1)AB}{\cal D}_A{\cal D}_B f$ in Eq.~(\ref{dgur}) vanishes because of ${\cal D}_A {\cal D}_B f 
\propto h_{AB}^{(0)}$ and the trace-free property of $h^{(1)AB}$. Noting that condition of (\ref{fA-cond}) 
is rearranged as ${\cal D}_A f_B+{\cal D}_B f_{A}=(2/3){\cal D}^C f_{C}h_{AB}^{(0)}$, we can show 
$\delta g_{AB}=O(r^{1/2})$. As a consequence, the transformation satisfying 
the conditions of Eqs.~(\ref{f0-cond}), (\ref{fA-cond}) and (\ref{f-cond}) does not 
disturb the boundary condition of Eqs.~(\ref{bc5}), 
(\ref{bc6}), (\ref{bc7}) and (\ref{bc8}) at leading order. 
By straightforward calculation, it can be shown that these transformations keep the metric
satisfying the boundary conditions at next-to-leading order.

Thus, asymptotic symmetry is generated by $f$ and $f^{A}$ satisfying Eqs.~(\ref{f0-cond}), 
(\ref{fA-cond}) and 
(\ref{f-cond}). The part of $f$ not proportional to $u$  generates a translation group. 
Since this part has only five degrees of freedom,
there is no supertranslation freedom unlike in four dimensions
\footnote{The fact that supertranslation freedom does not exist 
in higher dimensions is firstly pointed out in \cite{Hollands:2003ie} by using conformal 
method in even dimensions. We show that in five dimensions using the Bondi coordinates. }%
.
$f^{A}$ generates a Lorentz transformation group, so asymptotic symmetry at null infinity in five dimensional 
space-time is a Poincare group which is semi-direct of a five dimensional transformation group and Lorentz 
group. 

\section{summary and discussion}

In this paper, we define asymptotic flatness at null infinity in five dimensional space-time 
by using the Bondi coordinates. In the conformal embedding method, we cannot show the smoothness of 
asymptotic structure at null infinity because the gravitational waves behave like 
$\Omega^{3/2} \sim r^{-3/2}$ near 
null infinity and the regularity of gravitational fields at null infinity
is not guaranteed in general in five dimensions.  
On the other hand, in the Bondi coordinates, 
we can show the robustness of the asymptotic structure at null infinity which is defined by 
boundary conditions of Eqs.~(\ref{bc5}), (\ref{bc6}), (\ref{bc7}) and (\ref{bc8}). Solving Einstein equations under these boundary 
conditions, we find that total mass always decreases by gravitational waves as in four dimensions. 
In addition, we show that the asymptotic symmetry at null infinity would be a Poincare group in five dimensions. 

In four dimensions, asymptotic symmetry at null infinity is not a Poincare group. There are 
so called supertranslation freedoms, i.e., there are infinite dimensional translations. This supertranslation 
freedom comes from the freedom of the Bondi coordinates. In the Bondi coordinates, coordinate transformation
is described by the parameter $f$, $f^{A}$ in Eqs.~(\ref{xir}), (\ref{xiA}) and (\ref{xiu}), 
which corresponds to a translation and Lorentz transformation, respectively. 
If $f^{A}=0$ (pure translation) 
we may expect that $f$ has only four independent solutions. In general, however, there are conditions 
that $f$ should be a functions on a $2$-sphere and then 
$f$ has functional freedom, i.e., infinite dimensional degrees of freedom. 
This infinite dimensional set, which has an Abelian group structure, is called a 
supertranslational group. 

In four dimensional Minkowski space-time without any physical perturbations, the term $O(r)$ of $\delta g_{AB}$ 
should vanish, and this condition    
reduces infinite dimensional supertranslation to four dimensional translation.
In general, however, since gravitational waves contribute to $g_{AB}$ with $O(r)$
terms, which is the same order with metric variance, we cannot reduce supertranslation to translation. That is, we cannot distinguish the 
supertranslational ambiguity from gravitational waves. Thus the asymptotic symmetry at null infinity
in four dimensions is not a Poincare group. 

On the other hand, as shown in this paper, in five dimensions, there is no supertranslational
freedoms. This is because the behavior of gravitational waves in five dimensions is $1/r^{3/2}$ and 
this contributes to $g_{AB}$ with $O(r^{1/2})$ terms. 
In asymptotically flat five dimensional space-time, 
the term of $O(r)$ in $\delta g_{AB}$, which could be 
a contribution from the supertranslation,  
should vanish to maintain asymptotic flatness. 
This condition eliminates supertranslational freedom, 
and makes the asymptotic symmetry a Poincare group. 
Although we have only shown this feature in five dimensions, we expect that this feature would be common to 
the higher dimensional space-time in general, because gravitational waves in $d>4$ dimensions 
decay $1/r^{(d-2)/2}$ near null infinity faster than  supertranslation $O(1/r)$. 

In four dimensions, since there are supertranslations, we cannot choose a preferred rotational axis. 
The definition of angular momentum at null infinity does not have a precise meaning. 
%We cannot distinguish 
%supertranslations from the angular momentum loss by gravitational waves in angular momentum 
%change.   
Supertranslations and gravitational waves both contribute to angular momentum change, and we cannot 
distinguish one from another.
However, 
since there is no supertranslation in five dimensions, we can define total angular momentum at 
null infinity and observe the change of total angular momentum by gravitational waves. Furthermore, 
as we show the robustness of null infinity definition in five dimensions using the Bondi coordinates, it may 
be possible for us to redefine asymptotic flatness at null infinity using a conformal embedding method 
covariantly. This is left for future issue. 

Another remaining issue is the extension of our work to dimensions higher than seven. Using the 
Bondi coordinates, we have to go on step by step. It will be nice to have a systematic analysis 
for that. 
This is also our future work.

\acknowledgments
KT is supported by JSPS Grant-Aid for Scientific Research. 
NT and TS are partially supported by Grant-Aid for Scientific Research from Ministry of 
Education, Science, Sports and Culture of Japan (Nos.~2056381, 20540258, 21111006, 
and 19GS0219), the Japan-U.K. Research Cooperative Programs. 
This work is also supported by the Grant-in-Aid for the Global
COE Program ``The Next Generation of Physics, Spun from Universality and Emergenceh
from  the  Ministry  of  Education,  Culture,  Sports,  Science  and  Technology  (MEXT)  of
Japan.

%--- Acknowledgements ---%--- Acknowledgements ---%--- Acknowledgements ---%

\appendix

\section{Einstein equation}

We write the components of an Einstein equation explicitly in the expansion form with $1/r$. 
If for example, we expand function $C_{1}(u,r,x^{A})$ such that
%============<Equation>=============%
%
\begin{equation}
C_{1}\,=\,\frac{C_{11}(u,x^{A})}{r\sqrt{r}} +\frac{A(u,x^{A})}{r^{2}} +O(r^{-5/2}) ,
\end{equation} 
%
%===================================%
The Einstein equation $R_{\theta\theta}=0$ becomes
%============<Equation>=============%
%
\begin{equation}
-\frac{A_{,u}}{2r} + O(r^{-3/2}) = 0.
\end{equation}
%
%===================================%
That is, $\partial A(u,x^{A})/\partial u =0$. In Ref. \cite{Tanabe:2009xb}, it was shown that
from asymptotic flatness at spatial infinity, $A(u,x^{A})|_{u=-\infty}=0$. 
From this fact and $\partial A(u,x^{A})/\partial u =0$, we find that 
$A(u,x^{A})=0$ on null infinity. We can show in the same way that the 
$O(r^{-2})$ term in $C_{2}, D_{1}, D_{2}, D_{3}$ should also vanish. 

Thus, function $\mathbb{C}$ is expanded as following
%============<Equation>=============%
%
\begin{eqnarray}
& & C_{1}(u,r,x^{A})\,=\,\frac{C_{11}(u,x^{A})}{r\sqrt{r}} +\frac{C_{12}(u,x^{A})}{r^{2}\sqrt{r}}
+\frac{C_{13}(u,x^{A})}{r^{3}}+O(1/r^{7/2}) \\
& & C_{2}(u,r,x^{A})\,=\,\frac{C_{21}(u,x^{A})}{r\sqrt{r}} +\frac{C_{22}(u,x^{A})}{r^{2}\sqrt{r}}
+\frac{C_{23}(u,x^{A})}{r^{3}}+O(1/r^{7/2}) \\
& & D_{1}(u,r,x^{A})\,=\,\frac{D_{11}(u,x^{A})}{r\sqrt{r}} +\frac{D_{12}(u,x^{A})}{r^{2}\sqrt{r}}
+\frac{D_{13}(u,x^{A})}{r^{3}}+O(1/r^{7/2}) \\
& & D_{2}(u,r,x^{A})\,=\,\frac{D_{21}(u,x^{A})}{r\sqrt{r}} +\frac{D_{22}(u,x^{A})}{r^{2}\sqrt{r}}
+\frac{D_{23}(u,x^{A})}{r^{3}}+O(1/r^{7/2}) \\
& & D_{3}(u,r,x^{A})\,=\,\frac{D_{31}(u,x^{A})}{r\sqrt{r}} +\frac{D_{32}(u,x^{A})}{r^{2}\sqrt{r}}
+\frac{D_{33}(u,x^{A})}{r^{3}}+O(1/r^{7/2}).
\end{eqnarray}
%
%===================================%
Then we find from the Einstein equation that the other metric function in (\ref{metric}) should be 
expanded as follows 
%============<Equation>=============%
% 
\begin{eqnarray}
& & V\,=\,r^{2} +V_{1}(u,x^{A})\sqrt{r} +M(u,x^{A}) +O(1/r^{1/2}) \\
& & B\,=\,\frac{B_{1}(u,x^{A})}{r^{3}} + \frac{B_{2}(u,x^{A})}{r^{4}} +O(1/r^{9/2}) \\
& & U^{A} \,=\,\frac{U_{A1}(u,x^{A})}{r^{5/2}}+\frac{U_{A2}(u,x^{A})}{r^{7/2}}+\frac{U_{A3}(u,x^{A})}{r^{4}}
+O(1/r^{9/2}) .
\end{eqnarray}
%
%===================================%
Now we can write down each component of the Einstein equation as follows:
\vskip 5mm

\noindent $R_{rr}=0$:
%============<Equation>=============%
%
\begin{eqnarray}
&&-\frac{9}{8r^{5}}\left( 8B_{1} +C_{11}^{2} +C_{11}C_{21} +C_{21}^{2} +D_{11}^{2} +D_{21}^{2} +D_{31}^{2} \right) \notag\\
&&-\frac{3}{8r^{6}}\left( 32B_{2} +10C_{11}C_{12} +5C_{12}C_{21} +5C_{11}C_{22} +10C_{21}C_{22} \right. 
\notag \\
&& ~~~~~\left. +10D_{11}D_{12} +10D_{21}D_{22} +10D_{31}D_{32} \right) +O(r^{-13/2}) \,=\,0,
\end{eqnarray}
%
%===================================%
$R_{r\theta}=0$:
%=============<Equation>============%
%
\begin{eqnarray}
&&-\frac{3}{8r^{5/2}}\left( 2C_{11}\cot\theta -2C_{21}\cot\theta -4C_{11}\tan\theta -2C_{21}\tan\theta 
\right. \notag \\
&&~~~~~~~~~~~~~~~~~~~~~~~~~~~~-5U_{\theta 1} 
\left. + 2D_{21,\psi}\sec\theta +2D_{11,\phi}\csc\theta +2C_{11,\theta} \right) \notag\\
&&+\frac{1}{8r^{7/2}}\left( -10C_{12}\cot\theta +10C_{22}\cot\theta +20C_{12}\tan\theta 
+10C_{22}\tan\theta \right. \notag \\
&&~~~~~~~~~~~~~~~~~~~~~~~~~~~~
\left. +7U_{\theta 2}-10D_{22,\psi}\sec\theta -10D_{12,\phi}\csc\theta 
-10C_{12,\theta} \right) \notag \\
&&~~~ +O(r^{-4})\,=\,0,
\end{eqnarray}
%
%===================================%
$R_{r\phi}=0$:
\begin{eqnarray}
&&-\frac{3}{8r^{5/2}}\left( 4D_{11}\cos\theta -2D_{11}\sin\theta\tan\theta -5U_{\phi 1}\sin^{2}\theta 
\right. \notag \\
&&~~~~~~~~~~~~~~~~~~~~\left. +2D_{31,\psi}\tan\theta +2C_{21,\phi} +2D_{11,\theta}\sin\theta \right) \notag\\
&&+\frac{1}{8r^{7/2}}\left( -20D_{12}\cos\theta +10D_{12}\sin\theta\tan\theta +7U_{\phi 2}\sin^{2}\theta
\right. \notag \\ 
&&~~~~~~~~~~~~~~~~~~~~\left. -10D_{12,\theta}\sin\theta -10C_{22,\phi} -10D_{32,\psi}\tan\theta \right) 
\notag\\
%\notag\\
%&&-\frac{3}{8r^{4}}\left( -4C_{11}D_{11}\cos\theta -4C_{21}D_{11}\cos\theta +8D_{13}\cos\theta 
% -4D_{21}D_{31}\cos\theta \right. \notag\\
%&&~~+2C_{11}D_{11}\sin\theta\tan\theta +2C_{21}D_{11}\sin\theta\tan\theta -4D_{13}\sin\theta\tan\theta 
% +2D_{21}D_{31}\sin\theta\tan\theta \notag\\
%&&~~+2D_{31}C_{11,\psi}\tan\theta -2D_{21}D_{11,\psi}\tan\theta -2D_{11}D_{21,\psi}\tan\theta
% +2C_{11}D_{31,\psi}\tan\theta \notag
%\notag\\
%&&~~+4D_{33,\psi}\tan\theta -8B_{1,\phi} -2C_{11}C_{11,\phi} -C_{21}C_{11,\phi} -C_{11}C_{21,\phi} 
%\notag\\
%&&~~+4C_{23,\phi} -D_{11}D_{11,\phi} -2D_{21}D_{21,\phi} -6D_{31}D_{31,\phi} -2D_{11}C_{11,\theta}\sin\theta 
%\notag\\
%&&~~2D_{11}C_{21,\theta}\sin\theta -2C_{11}D_{11,\theta}\sin\theta -2C_{21}D_{11,\theta}\sin\theta 
% +4D_{13,\theta}\sin\theta 
%\notag\\
%&&\left. ~~-2D_{31}D_{21,\theta}\sin\theta -2D_{21}D_{31,\theta}\sin\theta \right) 
&&~~~~+ O(r^{-4})=0,
\end{eqnarray}
$R_{r\psi}=0$:
\begin{eqnarray}
&&\frac{3}{8r^{5/2}}(-2D_{21}\cot\theta\cos\theta +4D_{21}\sin\theta +5U_{\psi 1}\cos^{2}\theta\notag \\
&&~~~~~~~~~~~~~~~~~~ +2C_{11,\psi}
 +2C_{21,\psi} -2D_{31,\phi}\cot\theta -2D_{21,\theta}\cos\theta )\notag\\
&&+\frac{1}{8r^{7/2}}(-10D_{22}\cos\theta\cot\theta +20D_{22}\sin\theta +7U_{\psi 2}\cos^{2}\theta \notag\\
&&~~~~~~~~~~~~~~~~~~~+10C_{12,\psi} +10C_{22,\psi} -10D_{32,\phi}\cot\theta -10D_{22,\theta}\cos\theta ) 
\notag \\
%\notag\\
%&&+\frac{3}{8r^{4}}\left( 2C_{21}D_{21}\cos\theta\cot\theta +4D_{23}\cos\theta\cot\theta 
% -2D_{11}D_{31}\cos\theta\cot\theta -4C_{21}D_{21}\sin\theta \right. \notag\\
%&&~~-8D_{23}\sin\theta +4D_{11}D_{31}\sin\theta -8B_{1,\psi} -4C_{13,\psi} -C_{21}C_{11,\psi} \notag\\
%&&~~-2C_{21}C_{21,\psi} -4C_{23,\psi} +6D_{11}D_{11,\psi} +2D_{21}D_{21,\psi} +2D_{31}D_{31,\psi} \notag\\
%&&~~+2D_{31}C_{11,\phi}\cot\theta -2D_{21}D_{11,\phi}\cot\theta -2D_{11}D_{21,\phi}\cot\theta 
% +2C_{11}D_{31,\phi}\cot\theta \notag\\
%&&~~+4D_{33,\phi}\cot\theta +2D_{21}C_{21,\theta}\cos\theta -2D_{31}D_{11,\theta}\cos\theta 
% +2C_{21}D_{21,\theta}\cos\theta \notag\\
%&&~~\left. +2C_{21}D_{21,\theta}\cos\theta +4D_{23,\theta}\cos\theta -2D_{11}D_{31,\theta}\cos\theta 
%\right)  
&&~~~~~~~~+O(r^{-9/2})=0 ,
\end{eqnarray}
%
%===================================%
$R_{\theta\theta}=0$:
\begin{eqnarray}
&&\frac{1}{8r^{3/2}}\left( -13C_{11} -8U_{\theta 1}\cot\theta +8U_{\theta 1}\tan\theta -4V_{1} -8D_{21,\psi}\sec\theta\tan\theta 
\right. \notag \\
&&~~~~~~~~~~~~-8U_{\psi 1,\psi} -8U_{\phi 1,\phi} -4C_{11,\phi\phi}\csc^{2}\theta -4C_{11,\psi\psi}\sec^{2}
\theta \notag\\
&&~~~~~~~~~~~~+8D_{11,\phi}\cot\theta\csc\theta +4C_{11,\theta}\cot\theta -12C_{11,\theta}\tan\theta 
 \notag\\
&&~~~~~~~~~~~~\left.
-8C_{21,\theta}\tan\theta -12U_{\theta 1,\theta} +8D_{21,\theta\psi}\sec\theta +8D_{11,\theta\phi}\csc\theta  \right.\notag \\
&&~~~~~~~~~~~~~
\left.+4C_{11,\theta\theta}
 -8C_{21,\theta}\cot\theta -8C_{12,u}  \right) \notag \\
&&+\frac{3}{2r^{2}}\left(
C_{13,u} -D_{11}D_{11,u} -D_{21}D_{21,u}
\right) +O(r^{-5/2})=0,  
\end{eqnarray}
$R_{\theta\phi}=0$:
\begin{eqnarray}
&&\frac{1}{8r^{3/2}}\left( 3D_{11}\sin\theta -4D_{31,\psi} -4D_{31,\psi}\tan^{2}\theta 
 -4D_{11,\psi\psi}\sec\theta\tan\theta \right. \notag \\
&&~~ -4C_{11,\phi}\cot\theta -8C_{11,\phi}\tan\theta -4C_{21,\phi}\cot\theta -4C_{21,\phi}\tan\theta
 -2U_{\theta 1,\phi}\notag \\
&&~~+4D_{21,\phi\psi}\sec\theta -2U_{\phi 1,\theta}\sin^{2}\theta +4D_{31,\theta\psi}\tan\theta 
 +4C_{11,\theta\phi} \notag\\
&&~~\left. +4C_{21,\theta\phi} -8D_{12,u}\sin\theta \right) \notag \\
&&+\frac{3\sin\theta}{4r^{2}}\left( D_{11}C_{11,u} +D_{11}C_{21,u} +C_{11}D_{11,u} +C_{21}D_{11,u} \right. \notag\\
&&~~\left. -2D_{13,u} +D_{31}D_{21,u} +D_{21}D_{31,u} \right) +O(r^{-5/2}) =0, 
\end{eqnarray}
$R_{\theta\psi}=0$:
\begin{eqnarray}
&&\frac{1}{8r^{3/2}}( 3D_{21}\cos\theta +4C_{11,\psi}\cot\theta -4C_{21,\psi}\cot\theta
 -4C_{21,\psi}\tan\theta  \notag \\
&&~~~~~~~~-2U_{\theta 1,\psi} +4D_{31,\phi} 
+4D_{31,\phi}\cot^{2}\theta +4D_{11,\phi\psi}\csc\theta -4D_{21,\phi\phi}\cot\theta\csc\theta 
 \notag \\
&&~~~~~~~~-2U_{\psi 1,\theta}\cos^{2}\theta -4C_{21,\theta\psi} 
 +4D_{31,\theta\phi}\cot\theta -8D_{22,u}\cos\theta ) \notag\\
&&+\frac{3\cos\theta}{4r^{2}}\left( D_{21}C_{21,u} -D_{31}D_{11,u} +C_{21}D_{21,u}  
 +D_{23,u} -D_{11}D_{31,u} \right) +O(r^{-5/2})\,=\,0 ,
\end{eqnarray}
$R_{\phi\phi}=0$:
\begin{eqnarray}
&&\frac{1}{8r^{3/2}}\left( -16C_{11}\sin^{2}\theta +3C_{21}\sin^{2}\theta -12U_{\theta 1}\cos\theta\sin\theta
 +8U_{\theta 1}\sin^{2}\theta\tan\theta -4V_{1}\sin^{2}\theta  \right. \notag \\
&&~~+8D_{21,\psi}\sin\theta -8U_{\psi 1,\psi}\sin^{2}\theta -4C_{21,\psi\psi}\tan^{2}\theta
 +8D_{11,\phi}\cos\theta -8D_{11,\phi}\sin\theta\tan\theta \notag\\
&&~~-12U_{\phi 1,\phi}\sin^{2}\theta +8D_{31,\phi\psi}\tan\theta +4C_{21,\phi\phi} 
 +8C_{11,\theta}\cos\theta\sin\theta -4C_{21,\theta}\cos\theta\sin\theta \notag \\
&&~~+4C_{21,\theta}\sin^{2}\theta\tan\theta -8U_{\theta 1,\theta}\sin^{2}\theta +8D_{11,\theta\phi}\sin\theta 
 -4C_{21,\theta\theta}\sin^{2}\theta -8C_{22,u}\sin^{2}\theta \left.\right) \notag\\
&&-\frac{3\sin^{2}\theta}{2r^{2}}\left(
C_{23,u} -D_{11}D_{11,u} -D_{31}D_{31,u}\right) +O(r^{-5/2})=0  ,
\end{eqnarray}
$R_{\phi\psi}=0$:
\begin{eqnarray}
&&\frac{1}{8r^{3/2}}( 4D_{31}\cos^{2}\theta\cot\theta +11D_{31}\cos\theta\sin\theta 
 +4D_{31}\sin^{2}\theta\tan\theta +8D_{11,\psi}\cos\theta \notag \\
&&~~~~~~~+4D_{11,\psi}\sin\theta\tan\theta -2U_{\phi 1,\psi}\sin^{2}\theta -4D_{21,\phi}
\cos\theta\cot\theta -8D_{21,\phi}\sin\theta \notag \\ 
&&~~~~~~~ -2U_{\psi 1,\phi}\cos^{2}\theta -4C_{11,\phi\psi} -4D_{31,\theta}\cos^{2}\theta 
 +4D_{31,\theta}\sin^{2}\theta +4D_{11,\theta\psi}\sin\theta \notag \\
&&~~~~~~~ +4D_{21,\theta\phi}\cos\theta -D_{31,\theta\theta}\cos\theta\sin\theta 
 -8D_{32,u}\cos\theta\sin\theta  ) \notag \\
&&-\frac{3\cos\theta\sin\theta}{4r^{2}}\left( D_{31}C_{11,u} -D_{21}D_{11,u} -D_{11}D_{21,u} 
 +C_{11}D_{31,u} +D_{33,u}   \right) \notag \\
&&~~~~+O(r^{-5/2}) =0 ,
\end{eqnarray}
$R_{\psi\psi}=0$:
\begin{eqnarray}
&&\frac{1}{8r^{3/2}}\left( -19C_{11}\cos^{2}\theta -3C_{21}\cos^{2}\theta -8U_{\theta 1}\cos^{2}\theta\cot\theta
 -4V_{1}\cos^{2}\theta +12U_{\theta 1}\cos\theta\sin\theta \right. \notag \\
&&~~~~~+8D_{21,\psi}\cos\theta\cot\theta -8D_{21,\psi}\sin\theta 
-12U_{\psi 1,\psi}\cos^{2}\theta -4C_{11,\psi\psi} -4C_{21,\psi\psi}  
 \notag \\
&&~~~~~-8D_{11,\phi}\cos\theta -8U_{\phi 1,\phi}\cos^{2}\theta +8D_{31,\phi\psi}\cot\theta
+4C_{11,\phi\phi}\cot^{2}\theta +4C_{21,\phi\phi}\cot^{2}\theta 
 \notag\\
&&~~~~~+4C_{11,\theta}\cos^{2}\theta\cot\theta 
-12C_{11,\theta}\cos\theta\sin\theta +4C_{21,\theta}\cos^{2}\theta\cot\theta 
-4C_{21,\theta}\cos\theta\sin\theta  \notag \\
&&~~~~~~-8U_{\theta 1,\theta}\cos^{2}\theta+8D_{21,\theta\psi}\cos\theta 
 +8C_{11,\theta\theta}\cos^{2}\theta +8(C_{12,u}+C_{22,u})\cos^{2}\theta\left.\right) \notag \\
&&+\frac{3\cos^{2}\theta}{2r^{2}}\left( C_{13,u} +C_{23,u} -2D_{11}D_{11,u} -D_{21}D_{21,u}
 -D_{31}D_{31,u} \right) +O(r^{-5/2}) =0  ,
\end{eqnarray}
$R_{uu}=0$:
\begin{eqnarray}
&&\frac{1}{2r^{5/2}}( 3V_{1,u} +2(\cot\theta -\tan\theta)U_{\theta 1,u} +2U_{\theta 1,u\theta} 
 +2U_{\phi 1,u\phi} +2U_{\psi 1,u\psi} ) \notag \\
&&-\frac{1}{2r^{3}}\left( 3M_{,u} -(C_{11,u})^{2} -C_{11,u}C_{21,u} -(C_{21,u})^{2} -(D_{11,u})^{2}
 \right. \notag\\
&&~~~~~~~~~~~~~~~~~~~~~~~~~~~~~~~~~-(D_{21,u})^{2} -(D_{31,u})^{2}\left. \right) +O(r^{-7/2})=0,
\end{eqnarray}
$R_{u\theta}=0$:
%============<Equation>===============
%
\begin{eqnarray}
&&-\frac{1}{4r^{3/2}}( 2C_{11,u}\cot\theta -4C_{11,u}\tan\theta +2C_{21,u}\cot\theta 
-2C_{21,u}\tan\theta \notag \\
&&~~~~~~~~~~~~~~~~-5U_{\theta 1,u} +D_{21,u\psi}\sec\theta D_{11,u\phi}\csc\theta
+2C_{11,u\theta}  ) \notag \\
&&+\frac{1}{8r^{5/2}}\left( -U_{\theta 1} -8U_{\psi 1,\psi}\tan\theta -4U_{\theta 1,\psi\psi}\sec^{2}\theta
+8U_{\phi 1,\phi}\cot\theta  \right. \notag \\
&&~~~~~~-4U_{\theta 1,\phi\phi}\csc^{2}\theta -2V_{1,\theta} +4U_{\psi 1,\theta\psi} +4U_{\phi 1,\theta\phi} 
+4C_{12,u}\cot\theta \notag \\
&&~~~~~~-8C_{12,u}\tan\theta -4C_{22,u}\cot\theta -4C_{22,u}\tan\theta -14U_{\theta 2,u} \notag \\
&&~~~~~~~~~~~~~~+4D_{22,u}\sec\theta +4D_{12,u}\csc\theta +4C_{12,u}\left. \right) +O(r^{-4}) =0,
\end{eqnarray}
%
%=====================================
$R_{u\phi}=0$:
\begin{eqnarray}
&&\frac{1}{r^{3/2}}( 4D_{11,u}\cos\theta -2D_{11,u}\sin\theta\tan\theta -5U_{\phi 1,u}\sin^{2}\theta 
 +2D_{31,u\psi}\tan\theta \notag \\
 &&~~~~~~~~~~~~~~~~~~~~~~~~~~~~~~~~+2C_{21,u\phi} +2D_{11,u\theta}\sin\theta ) \notag \\
&&~~+\frac{1}{8r^{5/2}}( 15U_{\phi 1,\psi\psi}\sin^{2}\theta -4U_{\phi 1,\psi\psi}\tan^{2}\theta
 -4U_{\theta 1,\phi}\cot\theta -4U_{\theta 1,\phi}\tan\theta \notag \\
&&~~~~~~~-2V_{1,\phi} +4U_{\psi 1,\phi\psi} -12U_{\phi 1,\theta}\cos\theta\sin\theta 
 +4U_{\phi 3,\theta}\sin^{2}\theta\tan\theta \notag \\
&&~~~~~~~+4U_{\theta 1,\theta\phi} -4U_{\phi 3,\theta\theta}\sin^{2}\theta +D_{12,u\theta}\sin\theta    
 -4D_{12,u}\sin\theta\tan\theta -14U_{\phi 2,u}\sin^{2}\theta \notag \\
&&~~~~~~~~~~~~~~~~~~~~~+4D_{32,u\psi}\tan\theta 
+4C_{22,u\phi} +4D_{12,u\theta}\sin\theta ) 
%\notag \\
%&&+\frac{1}{4r^{2}}\left( -2V_{1,\phi} -2C_{11,\phi}C_{11,u} -C_{21,\phi}C_{21,u} 
% -4D_{11}C_{21,u}\cos\theta +2D_{11}C_{21,u}\sin\theta\tan\theta   \right. \notag \\
%&&~~-5U_{\phi 3}C_{21,u}\sin^{2}\theta -2D_{31,\psi}C_{21,u} -C_{11,\phi}C_{21,u} -2C_{21,\phi}C_{21,u}
% -2D_{11,\theta}C_{21,u}\sin\theta -4C_{11}D_{11,u}\cos\theta \notag \\
%&&~~+2C_{11}D_{11,u}\sin\theta\tan\theta -5U_{\theta 1}D_{11,u}\sin\theta -2D_{21,\psi}D_{11,u}\tan\theta 
% -4D_{11,\phi}D_{11,u} -2C_{11,\theta}D_{11,u}\sin\theta +4D_{13,u}\cos\theta \notag \\
%&&~~-2D_{12,u}\sin\theta\tan\theta -2D_{21,\phi}D_{21,u} -4D_{21}D_{31,u}\cos\theta 
% +2D_{21}D_{31,u}\sin\theta\tan\theta -5U_{\phi 3}D_{31,u}\cos\theta\sin\theta \notag\\
%&&~~+2C_{11,\psi}D_{31,u}\tan\theta +2C_{21,\psi}D_{31,u}\tan\theta -4D_{31,\phi}D_{31,u} 
% -2D_{21,\theta}D_{31,u}\sin\theta -5D_{11}U_{\theta 1,u}\sin\theta -5C_{21}U_{\phi 3,u}\sin^{2}\theta \notag\\
%+2C_{11}D_{31,u\psi}\tan\theta +2C_{21}D_{31,u\psi}\tan\theta \notag\\
%&&~~+2D_{33,u\psi}\tan\theta -2B_{1,u\phi} +2C_{23,u\phi} -2D_{11}D_{11,u\phi} -2D_{31}D_{31,u\phi} 
%-2D_{11}C_{21,u\theta}\sin\theta \notag \\
%&&~~+2D_{13,u\theta}\sin\theta -2D_{21}D_{31,u\theta}\sin\theta -2C_{11}D_{11,u\theta} 
% \left. \right) 
+O(r^{-3})=0 ,   
\end{eqnarray}
$R_{u\psi}=0$:
\begin{eqnarray}
&&\frac{1}{4r^{3/2}}( 2D_{21,u}\cos\theta\cot\theta -4D_{21,u}\sin\theta 
 -5U_{\psi 1,u}\cos^{2}\theta \notag \\
&&~~~~~~~~~~~~~~~~~~~~~-2C_{11,u\psi} -2C_{21,u\psi} +2D_{31,u\phi}\cot\theta 
+2D_{21,u\theta}\cos\theta ) \notag \\
&&+\frac{1}{8r^{5/2}}\left( 15U_{\psi 1}\cos^{2}\theta +4U_{\theta 1,\psi}\cot\theta +4U_{\theta 1,\psi}\tan\theta
 -2V_{1,\psi}  \right. \notag \\
&&~~~~~~~~~+4U_{\phi 3,\phi\psi} -4U_{\psi 1,\phi\phi}\cot^{2}\theta
-4U_{\psi 1,\psi}\cos^{2}\theta\cot\theta +12U_{\psi 1,\theta}\cos\theta\sin\theta 
\notag\\
&&~~~~~~~~~ +4U_{\theta 1,\theta\psi}-8D_{22,u}\sin\theta -14U_{\psi 2,u}\cos^{2}\theta 
-4C_{12,u\psi}-4C_{22,u\psi} \notag \\
&&~~~~~~~~~~~~~~~~~~~~~~~~~~~~~~~~~~~~~~~~~~~
+4D_{32,u\phi}\cot\theta +4D_{22,u\theta}\cos\theta \left.\right) \notag \\
%\notag\\
%&&+\frac{1}{4r^{3}}\left( -2M_{,\psi} +2D_{21}C_{11,u}\cos\theta\cot\theta -4D_{21}C_{11,u}\sin\theta 
% +5U_{\psi 1}C_{11,u}\cos^{2}\theta -2C_{11,\psi}C_{11,u} -C_{21,\psi}C_{11,u}\right. \notag\\
%&&~~+2D_{31,\phi}C_{11,u}\cot\theta +2D_{21,\theta}C_{11,u}\cos\theta +2D_{21}C_{21,u}\cos\theta\cot\theta
% -4D_{21}C_{21,u}\sin\theta +5U_{\psi 1}C_{21,u}\cos^{2}\theta -C_{11,\psi}C_{21,u} \notag\\
%&&~~-2C_{21,\psi}C_{21,u} +2D_{31,\phi}C_{21,u}\cot\theta +2D_{21,\theta}C_{21,u}\cos\theta 
% +2D_{11,\psi}D_{11,u} -2C_{11}D_{21,u}\cos\theta\cot\theta +4C_{11}D_{21,u}\sin\theta \notag \\
%&&~~-5U_{\theta 1}D_{21,u}\cos\theta -2D_{11,\phi}D_{21,u}\cot\theta -2C_{11,\theta}D_{21,u}\cos\theta 
% +2D_{23,u}\cos\theta\cot\theta -4D_{23,u}\sin\theta -2D_{11}D_{31,u}\cos\theta\cot\theta \notag\\
%&&~~+4D_{11}D_{31,u}\sin\theta -5U_{\phi 3}D_{31,u}\cos\theta\sin\theta -2C_{21,\phi}D_{31,u}\cot\theta 
% -2D_{11,\theta}D_{31,u}\cos\theta -5D_{21}U_{\theta 1,u}\cos\theta  \notag \\
%&&~~-5U_{\phi 3,u}D_{31}\cos\theta\sin\theta +5C_{11}U_{\psi 1,u}\cos^{2}\theta +5C_{21}U_{\psi 1,u}\cos^{2}\theta 
% -8U_{\psi 3,u}\cos^{2}\theta -2B_{1,u\psi} -2C_{13,u\psi} -2C_{23,u\psi}   \notag \\
%&&~~+4D_{11}D_{11,u\psi}+2D_{21}D_{21,u\psi}+2D_{31}D_{31,u} +2D_{31}C_{11,\phi}\cot\theta +2D_{31}C_{21,u\phi}\cot\theta 
% -2D_{11}D_{21,u\phi}\cot\theta   \notag\\
%&&~~-2C_{21}D_{31,u\phi}\cos\theta+2D_{33,u\phi}\cot\theta-2C_{11}D_{21,u\theta}\cos\theta 
% +2D_{23,u\theta}\cos\theta -2D_{11}D_{31,u\theta}\cos\theta  \left. \right) 
&&~~~~~+O(r^{-3}) =0.
\end{eqnarray}

\section{Derivation for Eqs. (\ref{divfA}) and (\ref{divF})}
%\section{Some details of calculations}

%\subsection{On commutativity between $\partial_u$ and ${\cal D}_A$ at null infinity}

%Let us exmine the commutativity between $\partial_u$ and ${\cal D}_A$. The direct 
%computation gives us 
%===========<Equation>==============%
%
%\begin{eqnarray}
%{\cal D}_A \partial_u X_B- \partial_u {\cal D}_A X_B
%=\partial_u {}^{(3)}\Gamma^C_{AB}X_C,
%\end{eqnarray}
%
%===================================% 
%where $\partial_u {}^{(3)}\Gamma^C_{AB} $ is the affine connection with respect to 
%$h_{AB}^{(0)}$ and $X_A$ is a vector on $S^3$. Since the right-hand side is roughly estimated 
%as $ \sim \partial_u \mathbb{C} =O(1/r^{3/2})$, we can confirm 
%the commutativity between $\partial_u$ and ${\cal D}_A$ at null infinity.

%\subsection{Derivation for Eqs. (\ref{divfA}) and (\ref{divF})}

Here we sketch the derivation for Eqs. (\ref{divfA}) and (\ref{divF}). 
We begin with 
%===========<Equation>==============%
%
\begin{eqnarray}
{\cal D}_B {\cal D}_AF =  {\cal D}_B {\cal D}_A {\cal D}_C f^C
& = & 
{\cal D}_B({\cal D}_C{\cal D}_A f^C-{}^{(3)}R_{AC}f^C) \nonumber \\
& = & -{\cal D}_B {\cal D}_C{\cal D}^C f_A-2{\cal D}_A{\cal D}_B \partial_u f  
-2{\cal D}_B f_A, \label{1}
\end{eqnarray}
%
%===================================% 
where ${}^{(3)}R_{AB}$ is the Ricci tensor with respect to $h_{AB}^{(0)}$. 
In the second and last line, we used the definition of Riemann tensor and Eq. (\ref{fA-cond}), respectively. 
We also used the fact that that Riemann tensor with respect to $h_{AB}^{(0)}$ is ${}^{(3)}R_{ABCD}=h_{AC}^{(0)}h_{BD}^{(0)}
-h_{AD}^{(0)}h_{BC}^{(0)}$. 
Using the definition of Riemann tensor two times, the first term in the last line of the right-hand side becomes 
%===========<Equation>==============%
%
\begin{eqnarray}
{\cal D}_B {\cal D}_C{\cal D}^C f_A
& = & {\cal D}_C {\cal D}_B{\cal D}^C f_A-2{\cal D}_Bf_A+h_{AB}^{(0)}{\cal D}_C f^C-{\cal D}_A f_B \nonumber \\
& = & {\cal D}^2{\cal D}_B f_A+2h_{AB}^{(0)}{\cal D}_C f^C-2({\cal D}_A f_B+{\cal D}_B f_A).
\end{eqnarray}
%
%===================================% 
Substituting this into Eq. (\ref{1}) and using the symmetry of indices $A$ and $B$, 
we have
%===========<Equation>==============%
%
\begin{eqnarray}
{\cal D}_B{\cal D}_A F=\frac{1}{2}\Biggl[ 
-{\cal D}^2({\cal D}_A f_B+{\cal D}_Bf_A)-4h_{AB}^{(0)}F
+2({\cal D}_A f_B+{\cal D}_Bf_A)-4{\cal D}_A {\cal D}_B \partial_u f
\Biggr].
\end{eqnarray}
%
%===================================% 
Then, using Eq. (\ref{fA-cond}) and $f_{,u}=-(1/3)F$, we obtain 
%===========<Equation>==============%
%
\begin{eqnarray}
{\cal D}_B{\cal D}_A F=-\frac{1}{3}h_{AB}^{(0)}{\cal D}^2F-\frac{4}{3}h_{AB}^{(0)}F+\frac{2}{3}
{\cal D}_A{\cal D}_B F. \label{2}
\end{eqnarray}
%
%===================================% 
The trace part implies 
%===========<Equation>==============%
%
\begin{eqnarray}
{\cal D}^2F+3F=0.
\end{eqnarray}
%
%===================================% 
Using this, Eq. (\ref{2}) becomes 
%===========<Equation>==============%
%
\begin{eqnarray}
{\cal D}_A {\cal D}_B F=\frac{1}{3}{\cal D}^2 F h_{AB}^{(0)}. 
\end{eqnarray}
%
%===================================% 

%---------   References   ---------%

%---------   References   ---------%

\end{document}